\documentclass[a4paper,fleqn,usenatbib]{mnras}
\usepackage{newtxtext,newtxmath}
\usepackage[T1]{fontenc}
\usepackage{ae,aecompl}
\usepackage{epsfig}
\usepackage{amsmath, amsfonts, epsfig, xspace}
\usepackage{natbib}
\usepackage{rotating}
\usepackage{graphicx}
\usepackage{caption}
\usepackage{longtable}
\usepackage{multicol}
\usepackage{booktabs}
\usepackage{amsmath}
\usepackage{rotating}
\usepackage{booktabs}
\usepackage{changepage}
\usepackage{scalerel}
\usepackage{float}
\usepackage[normalem]{ulem}
\usepackage{graphicx}
\usepackage{dcolumn}
\usepackage{bm}
\usepackage{color}
\usepackage{array}
\usepackage{multirow}
\usepackage{amsmath}
\usepackage{bm}
\usepackage{subfigure}
\usepackage{newtxtext,newtxmath}
\usepackage{orcidlink}
\title[A cosmological distance measure using radio-loud quasars]{A cosmological distance measure using radio-loud quasars}

\author[L. Huang, Z. X. Chang]{L. Huang \orcidlink{0000-0003-4545-7066} $^{1,3}$\thanks{E-mail:
huanglong20122021@163.com}, Z. X. Chang$^{2}$\\
$^{1}$College of Science, Jiujiang University, Jiujiang 332000, People's Republic of China.\\
$^{2}$College of Mathematical and Physical Sciences, HanDan University, Handan 056000, People's Republic of China.\\
$^{3}$Key Laboratory of Functional Microscale Materials in Jiangxi Province, Jiujiang 332000, People's Republic of China.\\}

\begin{document}
\date{Accepted ---. Received ---; in original form ---}

\pagerange{\pageref{firstpage}--\pageref{lastpage}} \pubyear{2021}

\maketitle

\label{firstpage}

\begin{abstract}

We use the X-ray luminosity relation of radio-loud quasars (RLQs) to measure these luminosity distances as well as estimate cosmological parameters. We adopt four parametric models of X-ray luminosity to test luminosity correlation for RLQs and radio-intermediate quasars (RIQs) and give these cosmological distances. By Bayesian information criterion (BIC), the data suggest that the luminosity relation ${L_X} \propto L_{UV}^{{\gamma _{uv}}}L_{Radio}^{\gamma _{radio}'}$ for RLQs has better goodness of fit, relative to other models, which can be interpreted as this relation being preferred for RLQs. Meanwhile, we compare the results from flat-spectrum radio-loud quasars (FSRLQs) and steep-spectrum radio-loud quasars (SSRLQs), which indicate that their luminosity correlations are not exactly the same. We also consider dividing the RLQs sample into various redshift bins, which can be used to check if the X-ray luminosity relation depends on the redshift. Finally, we apply a combination of RLQs and SNla Pantheon to verify the nature of dark energy concerning whether or not its density deviates from the constant, and give the statistical results.

\end{abstract}

\begin{keywords}
quasars: general--galaxies: high-redshift--cosmology: observations--cosmology: distance scale--cosmology: dark energy
\end{keywords}

\section{Introduction}

With the rapid development of the economy and astronomical technologies, more and more observational data can be used for cosmological distance measurement. Such as SNe la \citep{Riess1998,Amanullah2010,Betoule2014,Scolnic2018}, quasars \citep{Risaliti2015}, GRB \citep{Khadka2021}, gravitational wave \citep{Chen2018,Fu2019} and so on \citep{Jimenez2003}. On the other hand, these data can also be applied for the test of the Cosmological Principle \citep{Secrest2021}. We consider using radio-loud quasars (RLQs) with luminosity correlation to measure these cosmological luminosity distances.

Radio-loud quasars are one type of quasars, which are categorized by their radio luminosity. Since the discovery of quasars, it has been known that some of them have strong radio emissions and others do not. Strittmatteret al.\citet{Strittmatter1980} showed that a dichotomy can be used in the distribution of the radio luminosity of quasars, Kellermann et al.\citet{Kellermann1989} also defined different quasars based on the ratio of monochromatic luminosities. RLQs are often defined by a radio-loudness parameter satisfying $logR >1$ for the subset of quasars, where R is the ratio of monochromatic luminosities (with units of $erg{\kern 1pt} {\kern 1pt} {\kern 1pt} {s^{ - 1}}{\kern 1pt} {\kern 1pt} H{z^{ - 1}}$) measured at (rest-frame) $5 GHz$ and 2500 {\AA}  \citep{Stocke1992,Kellermann1994}. Radio-quiet quasars (RQQs) must minimally satisfy $logR< 1$ and often are found to have $logR < 0$. This is the traditional division for quasars. The physical reason for the difference between RLQs and RLQs populations remains elusive, several explanations involve the physical origin of radio emission \citep{Laor2008, Panessa2019}, black hole masses \citep{Lacy2001}, accretion rates \citep{Sikora2007}, and/or spins \citep{Garofalo2010}, which results that a universal function such as the traditional division may not be interpreted by the exact same physical mechanism \citep{Balokovic2012}, but such classical dichotomy is still relevant.

The X-ray properties of RLQs are different from those of RQQs. RLQs are generally more X-ray luminous than RQQs of matched optical/UV luminosity \citep{Willott2001,Laor2000,Meier2001,Metcalf2006,Shankar2010,Garofalo2010,Browne1987,Worrall1987}. The X-ray luminosity in RLQs may be correlated with optical/UV and radio luminosity, which can be verified by parameterization methods \citep{Tananbaum1983, Worrall1987, Miller2010, Zhu2020, Browne1987}. The luminosity correlation might indicate that X-ray emission is not merely created by Compton upscattering of disk photons occurring in a hot “corona”, but also powered directly or indirectly by the radio jet \citep{Evans2006, Hardcastle2009, Miller2010}, which makes RLQs regarded as standard candles and can be used for cosmological distance measurement.

Hence, we combine the parameterized model of X-ray luminosity with X-ray, optical/UV, and radio flux at (rest-frame)5 GHz,  2500 {\AA}, and 2kev to test luminosity correlation as well as obtain luminosity distance. On the other hand, Flat-spectrum radio-loud quasars (FSRLQs) and steep-spectrum radio-loud quasars (SSRLQs) are the two categories of RLQs \citep{Urry1995,Barthel1989,Padovani1992}, which can be defined by ${\alpha _r} >  - 0.5$ for FSRLQs, and ${\alpha _r} \le  - 0.5$ for SSRLQs. FSRQs and SSRQs might involve different mechanisms which can be used to explain their X-ray data.  Therefore, we consider applying various models to FSRLQs and FSRQs and seek optimal models.

In addition, Worrall et al have used RLQs to check whether luminosity correlations are redshift dependent \citep{Miller2010}. Hence, we also consider dividing the RLQs sample into various redshift bins, which can be used to constrain model parameters and examine whether or not luminosity relations depend on redshift.

In Section \ref{Sec:2} of this paper, we introduce the source of data used, including the flux of RLQs and RIQs in the radio, optical/UV, and X-ray wavebands. In Section \ref{Sec:3}, we employ four parametric models to test the X-ray luminosity correlation of RLQs and RIQs, which include X-ray luminosity as a sole function of optical/UV luminosity and as a joint function of optical/UV and radio luminosity. In Section \ref{Sec:4}, we compare and analyze four different models by using the Bayesian information criterion (BIC), meanwhile obtaining cosmological luminosity distance. Furthermore, we consider dividing the RLQs sample into various redshift bins, which can be used for checking if the X-ray luminosity relation depends on the redshift. In Section \ref{Sec:5}, we apply a combination of RLQs and SNla Pantheon to reconstruct the dark energy equation of state $w(z)$, which can be used to test the nature of dark energy concerning whether or not its density deviates from the constant. In Section \ref{Sec:6}, we summary the paper.

\section[ data used]{ data used} \label{Sec:2}

Radio (e.g., Faint Images  of the Radio Sky) surveys \citep{Helfand2015}, modern optical (e.g., Sloan Digital Sky Survey) \citep{Lyke2020,Paris2018,Alam2015,Richards2002},  and archival X-ray data from Chandra \citep{Evans2010}, XMM-Newton \citep{Traulsen2019}, and ROSAT \citep{Sun2012} provide large amount of quasars data, which can be used to investigate luminosity correlation for quasars. \citet{Miller2010} obtained 654 sample including RIQs and RLQs, which are selected from SDSS/FIRST observations and high quality X-ray coverage from Chandra, XMM-Newton, or ROSAT \citep{Suchkov2006,Jester2006,Green2009,Young2009,Schneider2007,Richards2009}. \citet{Zhu2020} further selected new RLQs using the Sloan Digital Sky Survey, and the radio data are from the Faint Images  of the Radio Sky at  Twenty-Centimeters, and the NRAO VLA Sky Survey \citep{Helfand2015,Paris2017,Banfield2015,York2000,Becker1995,Condon1998}. The X-ray luminosities can be obtaied from Chandra and XMM Newton observations \citep{Evans2010,Rosen2016}.

Therefore, we consider to use a combined data from  \citet{Miller2010} and \citet{Zhu2020}. The observed $1.4GHz$ flux is utilized to calculate rest-frame radio flux at 5GHz (${F_{radio}}$) by assuming ${\alpha _r} =  - 0.5$. In the same way, the i-band apparent magnitude (${m_i}$) can be used  for  calculating rest-frame optical/UV flux at 2500 {\AA}  (${F_{UV}}$), where an optical spectral index of ${\alpha _o} =  - 0.5$ are assumed for the K-correction \citep{Richards2006}. Count rates were also converted to the unabsorbed flux density at observed-frame 2keV using PIMMs, where a  specifying Galactic column density and a power-law spectrum $\Gamma  = 1.5$ are considered \citep{Miller2010}, which can be used to determine bandpass-corrected rest-frame 2 keV flux.

Meanwhile, a part of RIQs and RLQs from FIRST measured flux densities at $1.4GHz$ can not directly obtain their radio slopes, \citet{Zhu2020} matched those quasars with other radio surveys, the latter can provide radio fluxes at another wavelength, which can be applied to calculate radio slops. Other radio surveys include the Green Bank 6-cm (GB6) Radio Source Catalog at $4.85 GHz$ \citep{Gregory1996}, Westerbork Northern Sky Survey at $325 MHz$ \citep{Rengelink1997}, TGSS Alternative Data Release at $150 MHz$ \citep{Intema2017},  LoTSS DR1  at $144 MHz$ \citep{Shimwell2019}, and the Very Large Array Sky Survey (VLASS) at $3 GHz$ \citep{Hancock2012,Hancock2018}.  \citet{Zhu2020} also gather their multi-band radio flux densities from the NED\footnote{http://nedwww.ipac.caltech.edu/} or VizieR For RLQs that are not matched with those radio surveys.

In this paper, we only consider RLQs with $logR>2$, and RIQs satisfy $1 \le {\rm{logR}} \le {\rm{2}}$. Meanwhile, FSRLQs are given by ${\alpha _r} >  - 0.5$ , and SSRLQs have ${\alpha _r} \le  - 0.5$ . We adopt parametric methods to test their luminosity correlation.

\section{PARAMETERIZING THE X-RAY LUMINOSITY OF RLQs and RIQs}\label{Sec:3}

We use different parameterization methods to test luminosity correlation, which involve different physical mechanisms. The most common form is \citep{Tananbaum1983,Worrall1987,Miller2010,Zhu2020}
 \begin{equation}\label{eq1}
\begin{array}{l}
Model\\
I:\log {L_X} = \alpha  + {\gamma _{uv}}\log {L_{UV}} + \gamma _{radio}'\log {L_{Radio}},
\end{array}
\end{equation}

The above equation can become the relation ${L_X} \propto L_{UV}^{{\gamma _{uv}}}L_{Radio}^{\gamma _{radio}'}$. Using formula $L = 4\pi {D_L}^2F$ in it, we get
\begin{equation}\label{eq2}
\begin{array}{l}
\log {F_X} = \Phi ({F_{UV}},{F_{radio}},{D_L})\\
{\kern 1pt} {\kern 1pt} {\kern 1pt} {\kern 1pt} {\kern 1pt} {\kern 1pt} {\kern 1pt} {\kern 1pt} {\kern 1pt} {\kern 1pt} {\kern 1pt} {\kern 1pt} {\kern 1pt} {\kern 1pt} {\kern 1pt} {\kern 1pt} {\kern 1pt} {\kern 1pt} {\kern 1pt} {\kern 1pt} {\kern 1pt} {\kern 1pt} {\kern 1pt} {\kern 1pt} {\kern 1pt} {\kern 1pt} {\kern 1pt} {\kern 1pt} {\kern 1pt} {\kern 1pt} {\kern 1pt} {\kern 1pt} {\kern 1pt} {\kern 1pt}  = \alpha  + {\gamma _{uv}}\log {F_{UV}} + \gamma _{radio}'\log {F_{radio}}\\
{\kern 1pt} {\kern 1pt} {\kern 1pt} {\kern 1pt} {\kern 1pt} {\kern 1pt} {\kern 1pt} {\kern 1pt} {\kern 1pt} {\kern 1pt} {\kern 1pt} {\kern 1pt} {\kern 1pt} {\kern 1pt} {\kern 1pt} {\kern 1pt} {\kern 1pt} {\kern 1pt} {\kern 1pt} {\kern 1pt} {\kern 1pt} {\kern 1pt} {\kern 1pt} {\kern 1pt} {\kern 1pt} {\kern 1pt} {\kern 1pt} {\kern 1pt} {\kern 1pt} {\kern 1pt} {\kern 1pt} {\kern 1pt} {\kern 1pt} {\kern 1pt}  + ({\gamma _{uv}} + \gamma _{radio}' - 1)\log (4\pi {D_L}^2),
\end{array}
\end{equation}
where ${F_{X}}$, ${F_{UV}}$ and ${F_{radio}}$ are measured at (rest-frame) $2 keV$, 2500 {\AA}  and $5 GHz$, $D_L$ is the luminosity distance, which can be calculated by integral formula of $D_L-z$ relation. Thus equation (\ref{eq2}) can be used to test  X-ray luminosity correlation for RLQs and RIQs.

We can aslo consider other models, including \citep{Miller2010,Zhu2020,Browne1987}
 \begin{equation}\label{eq3}
II:{\kern 1pt} {\kern 1pt} {\kern 1pt} {\kern 1pt} {\kern 1pt} {\kern 1pt} \log {L_X} = \log (AL_{UV}^{{\gamma _{uv}}} + BL_{Radio}^{{\gamma _{radio}}}),
\end{equation}

 \begin{equation}\label{eq4}
III:{\kern 1pt} {\kern 1pt} {\kern 1pt} {\kern 1pt} {\kern 1pt} {\kern 1pt} \log {L_X} = \log (AL_{UV}^{{\gamma _{uv}}}L_{Radio}^{\gamma _{radio}'} + BL_{Radio}^{{\gamma _{radio}}}),
\end{equation}

Model II and Model III can become the relation ${L_X} = AL_{UV}^{{\gamma _{uv}}} + BL_{Radio}^{{\gamma _{radio}}}$ and ${L_X} = AL_{UV}^{{\gamma _{uv}}}L_{Radio}^{\gamma _{radio}'} + BL_{Radio}^{{\gamma _{radio}}}$. The above three models concern that X-ray luminosity is related to both optical/UV luminosity and radio luminosity. The last model is considered that X-ray luminosity is only correlated with optical/UV luminosity, and its parametric form is \citep{Miller2010,Bisogni2021}
 \begin{equation}\label{eq5}
IV:{\kern 1pt} {\kern 1pt} {\kern 1pt} {\kern 1pt} {\kern 1pt} {\kern 1pt} \log {L_X} = \alpha  + {\gamma _{uv}}\log {L_{UV}},
\end{equation}

Similarly, from Equations (\ref{eq3}),(\ref{eq4}) and (\ref{eq5}), we can get X-ray flux ${F_{X}}$ as the function of ${F_{UV}}$, ${F_{radio}}$ and $D_L$, which can be used to verify  X-ray luminosity relation.

\section{Models constrains from RLQs and RIQs}\label{Sec:4}
\subsection{Fitting the Models to RLQs and RIQs}

We fit parametric models by minimizing a likelihood function consisting of a modified ${\chi ^2}$ function based on MCMC, allowing for an intrinsic dispersion $\sigma $ \citep{Risaliti2015}
\begin{small}
 \begin{equation}\label{eq6}
 - 2\ln L = \sum\limits_{i = 1}^N {\left\{ {\frac{{{{[\log {{({F_X})}_i} - \Phi {{({F_{UV}},{F_{radio}},{D_L})}_i}]}^2}}}{{s_i^2}} + \ln (2\pi s_i^2)} \right\},}
\end{equation}
\end{small}
where $\Phi ({F_{UV}},{F_{radio}},{D_L})$ is given by equation (\ref{eq2}), and ${s_i} = \sigma $, $\sigma$ is the intrinsic dispersion, which can be fitted as a free parameter. Also, $\sigma$ is much larger than measurement error in most cases.

The parameters $\alpha $ and Hubble constant ${H_0}$ are degenerate when fitting equation (\ref{eq2}), we fix ${H_0} = 70{\kern 1pt} {\kern 1pt} km{\kern 1pt} {\kern 1pt} {\kern 1pt} {s^{ - 1}}{\kern 1pt} Mp{c^{ - 1}}$ \citep{Reid2019,Aghanim2020}. If we want to better test X-ray luminosity relation and further seek optimal model, we should not fix ${\Omega _m}$. Therefore, we fit four models to RIQs and RLQs without fixing ${\Omega _m}$, and select best model.

Meanwhile, we measure the distance modulus for FSRLQs and SSRlQs. For Model I, Equation (\ref{eq2}) gives distance modulus as
 \begin{equation}\label{eq7}
DM = \frac{{5[\log {F_X} - {\gamma _{uv}}\log {F_{UV}} - \gamma _{radio}'{F_{radio}} - {\alpha '}]}}{{2({\gamma _{uv}} + \gamma _{radio}' - 1)}},
\end{equation}
where ${\alpha '} = \alpha  + ({\gamma _{uv}} + \gamma _{radio}' - 1)\log (4\pi )$.  The error is
 \begin{equation}\label{eq8}
{\sigma _{DM}} = \frac{{5\sigma }}{{2({\gamma _{uv}} + \gamma _{radio}' - 1)}}.
\end{equation}

In the same way, for model IV, the distance modulus is
 \begin{equation}\label{eq9}
DM = \frac{{5[\log {F_X} - {\gamma _{uv}}\log {F_{UV}} - \alpha ']}}{{2({\gamma _{uv}} - 1)}}.
\end{equation}
and its error
 \begin{equation}\label{eq10}
{\sigma _{DM}} = \frac{{5\sigma }}{{2({\gamma _{uv}} - 1)}},
\end{equation}

For Model II and Model III, distance modulus can be calculated by numerical solution of equations (\ref{eq3}) and (\ref{eq4}).

\begin{figure*}
\centering
\includegraphics[scale=0.9]{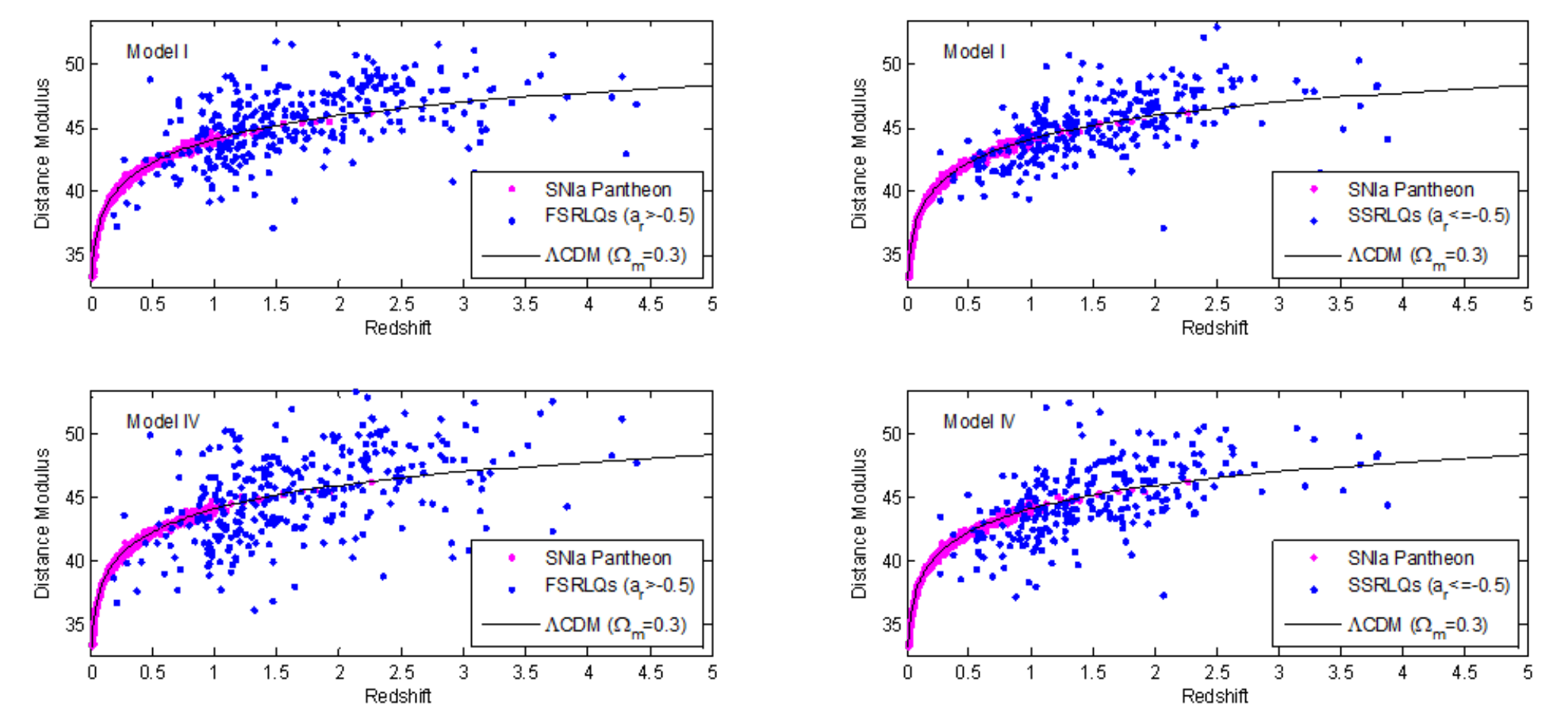}
\caption{Distance modulus from fitting Model I and Model IV with a prior flat $\Lambda CDM$ model to FSRLQs and SSRLQs. Blue points are FSRLQs and SSRLQs distance modulus, pink points are SNla from Pantheon sample. The blue line shows a flat $\Lambda CDM$ model fit with ${\Omega _m} = 0.3$.}
\label{fig:1}
\end{figure*}

\begin{figure*}
\centering
\includegraphics[width=\linewidth,scale=1.00]{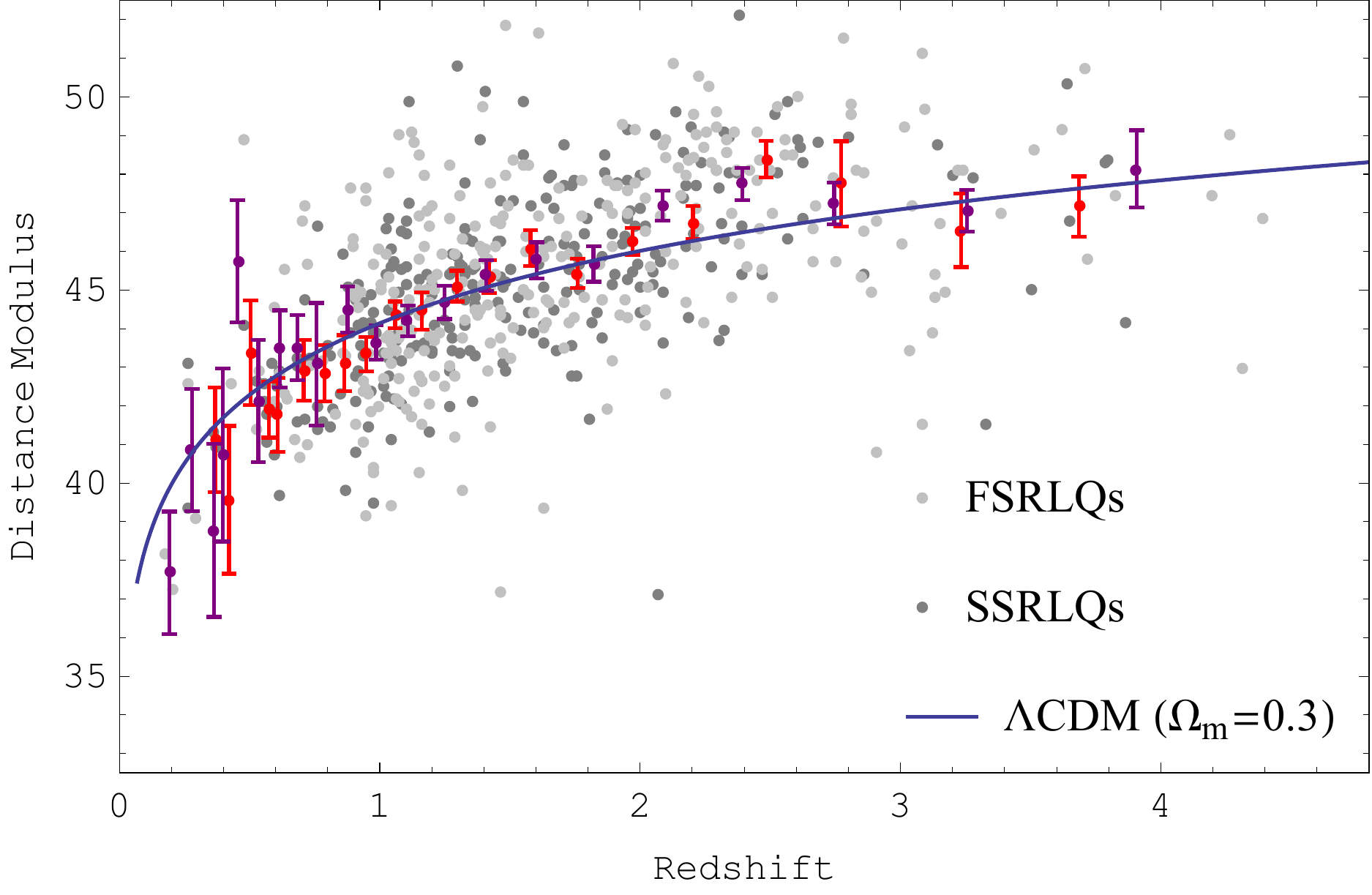}
\caption{FSRLQs (light grey points) and SSRLQs distance modulus (grey points) from a fit of Model I when assuming $\Lambda CDM$ cosmology. The purple and red points are FSRLQs and SSRLQs averages in small redshift bins. The blue line shows a flat
$\Lambda CDM$ model fit with ${\Omega _m} = 0.3$.}
\label{fig:2}
\end{figure*}

We adopt maximum likelihood function (equation (\ref{eq6})) based on MCMC to constrain four models, the fit results are shown in table \ref{tab:1}. Meanwhile, we compare FSRLQS and SSRLQS distance modulus from model I and model IV, which are shown in fig \ref{fig:1}. Fig \ref{fig:2} shows FSRLQs and SSRLQs distance modulus from a fit of Model I when assuming $\Lambda CDM$ cosmology, and their averages in small redshift bins.

\begin{table*}
\setlength{\tabcolsep}{0.5mm}
 \centering
  \caption{ Model fitting results for RIQs and RLQs }
  \label{tab:1}
  \vspace{0.3cm}
  \begin{tabular}{@{}ccccccccccc@{}}
  %&$q_0$&$j_0$&${t_0}{H_0}{\kern 1pt} ({t_0} = {T_0},{\eta _{{T_0}}})$&$\chi _{\min }^2$
  \hline
  Model&Sample&$\alpha $&$A $&${\gamma _{uv}} $&$B$&${\gamma _{radio}}$&$\gamma _{radio}'$&$\sigma $&${\Omega _m}$&$- 2In{L_{\max }}/N$\\
    \uppercase\expandafter{\romannumeral1}
   &RIQs&4.99$\pm$0.67 & $-$   &0.454$\pm$0.016&$-$&$-$&   0.245$\pm$0.01 & 0.369$\pm$0.02 &0.162$\pm$0.042 &119$/$144\\
   &FSRLQs&5.24$\pm$0.31 & $-$   &0.383$\pm$0.007&$-$&$-$&   0.301$\pm$0.009 & 0.282$\pm$0.012 &0.09$\pm$0.013 &95$/$305\\
   &SSRLQs&6.07$\pm$0.232 & $-$   &0.4$\pm$0.012&$-$&$-$&   0.259$\pm$0.008 & 0.261$\pm$0.01 &0.136$\pm$0.075 &32$/$266\\
   &RLQs&5.29$\pm$0.108 & $-$   &0.416$\pm$0.006&$-$&$-$&   0.27$\pm$0.006 & 0.276$\pm$0.008 &0.088$\pm$0.037 &147$/$571\\
    &RLQs+RIQs&5.33$\pm$0.176 & $-$   &0.456$\pm$0.007&$-$&$-$&   0.231$\pm$0.004 & 0.298$\pm$0.009 &0.175$\pm$0.029 &290$/$715\\
   \hline

    \uppercase\expandafter{\romannumeral2}
   &RIQs&$-$& 6.69$\pm$0.546 & 0.65$\pm$0.019&6$\pm$0.338   &0.626$\pm$0.007& $-$&  0.371$\pm$0.024 & 0.2$\pm$07  &118$/$144\\
   &FSRLQs&$-$& 5.93$\pm$0.284 & 0.68$\pm$0.009&4.37$\pm$0.282   &0.66$\pm$0.008& $-$&  0.29$\pm$0.012 & 0.11$\pm$0.063  &97$/$305\\
   &SSRLQs&$-$& 5.51$\pm$0.154 & 0.7$\pm$0.005&6.06$\pm$0.188   &0.61$\pm$0.005& $-$&  0.26$\pm$0.012 & 0.069$\pm$0.109  &35$/$266\\
   &RLQs&$-$& 5.71$\pm$0.18 & 0.69$\pm$0.006&6$\pm$0.207   &0.61$\pm$0.005& $-$&  0.278$\pm$0.008 & 0.084$\pm$0.083  &155$/$571\\
    &RLQs+RIQs&$-$& 5.99$\pm$0.146 & 0.6$\pm$0.005&6.02$\pm$0.09   &0.61$\pm$0.002& $-$&  0.3$\pm$0.007 & 0.22$\pm$0.09  &302$/$715\\
   \hline

    \uppercase\expandafter{\romannumeral3}
   &RIQs&$-$ & 7.14$\pm$0.557   &0.39$\pm$0.021&3.76$\pm$0.3&0.65(fixed)&   0.235$\pm$0.008 & 0.375$\pm$0.023 &0.44$\pm$0.08 &118$/$144\\
   &FSRLQs&$-$ & 4.62$\pm$0.322   &0.43$\pm$0.014&2.65$\pm$0.312&0.65(fixed)&   0.276$\pm$0.014 & 0.28$\pm$0.011 &0.085$\pm$0.03 &94$/$305\\
   &SSRLQs&$-$ & 6.26$\pm$0.292   &0.42$\pm$0.01&3.38$\pm$0.353&0.65(fixed)&   0.23$\pm$0.012 & 0.26$\pm$0.011 &0.105$\pm$0.05 &32$/$266\\
   &RLQs&$-$ & 5.45$\pm$0.373   &0.43$\pm$0.007&2.53$\pm$0.637&0.65(fixed)&   0.25$\pm$0.008 & 0.276$\pm$0.007 &0.062$\pm$0.034 &145$/$571\\
    &RLQs+RIQs&$-$ & 5.43$\pm$0.275   &0.46$\pm$0.006&3.28$\pm$0.393&0.65(fixed)&   0.225$\pm$0.008 & 0.3$\pm$0.008 &0.1$\pm$0.027 &289$/$715\\
   \hline

   \uppercase\expandafter{\romannumeral4}
  &RIQs&5.78$\pm$0.225 & $-$   &0.687$\pm$0.0076&$-$&$-$&   $-$ & 0.37$\pm$0.02 &0.35$\pm$0.088 &120$/$144\\
   &FSRLQs&4.38$\pm$0.351 & $-$   &0.741$\pm$0.012&$-$&$-$&   $-$ & 0.316$\pm$0.012 &0.06$\pm$0.034 &158$/$305\\
   &SSRLQs&5.47$\pm$0.311 & $-$   &0.703$\pm$0.01&$-$&$-$&   $-$ & 0.289$\pm$0.012 &0.056$\pm$0.041 &79$/$266\\
   &RLQs&5.96$\pm$0.32 & $-$   &0.704$\pm$0.01&$-$&$-$&   $-$ & 0.3$\pm$0.009 &0.06$\pm$0.027 &248$/$571\\
    &RLQs+RIQs&5.71$\pm$0.247 & $-$   &0.695$\pm$0.008&$-$&$-$&   $-$ & 0.33$\pm$0.008 &0.226$\pm$0.037 &440$/$715\\
   \hline

\end{tabular}
\end{table*}

\begin{table*}
%\footnotesize
\setlength{\tabcolsep}{0.5mm}
 \centering
  \caption{Fit results on model parameters for a combination of SNla and RLQs}
  \label{tab:2}
  \vspace{0.3cm}
  \begin{tabular}{@{}cccccccccccc@{}}
  %&$q_0$&$j_0$&${t_0}{H_0}{\kern 1pt} ({t_0} = {T_0},{\eta _{{T_0}}})$&$\chi _{\min }^2$
  \hline
&$Sample $&$\alpha $&${\gamma _{uv}} $&$\gamma _{radio}'$&$\sigma $&${\Omega _m}$&${w_0}$&${w_\alpha }$&$\chi _{Total}^2$/N\\
   $\Lambda CDM$
   &$SN+RLQs$ & 6.28$\pm$0.259   & 0.43$\pm$0.007& 0.225$\pm$0.01&   0.278$\pm$0.007 & 0.275$\pm$0.007 &$-$&$-$ &1151.1/1619\\
&$SN+FSRLQs$ & 4.13$\pm$0.526   & 0.475$\pm$0.014& 0.25$\pm$0.013&   0.285$\pm$0.012 & 0.276$\pm$0.007 &$-$&$-$ &1097/1353\\
&$SN+SSRLQs$ & 6.57$\pm$0.406   & 0.4$\pm$0.012& 0.24$\pm$0.008&   0.26$\pm$0.012 & 0.275$\pm$0.008 &$-$&$-$ &1036.8/1314\\

   \hline
      ${w_0}{w_a}CDM$
   &$SN+RLQs$ & 5.42$\pm$0.259   & 0.45$\pm$0.009& 0.23$\pm$0.006&   0.28$\pm$0.008 & 0.33$\pm$0.0017 &-1.168$\pm$0.047&0.232$\pm$0.517 &1147.7/1619\\
&$SN+FSRLQs$ & 5.35$\pm$0.17   & 0.41$\pm$0.014& 0.274$\pm$0.011&   0.286$\pm$0.011 & 0.3136$\pm$0.035 &-1.158$\pm$0.062&0.578$\pm$0.478&1091.3/1353\\
&$SN+SSRLQs$ & 6.8$\pm$0.463   & 0.42$\pm$0.01& 0.215$\pm$0.0018&   0.26$\pm$0.012 & 0.29$\pm$0.033 &-1.11$\pm$0.07&0.63$\pm$0.358&1032/1314\\

   \hline

\end{tabular}
\end{table*}

\subsection{Models analysis and comparison}

We use BIC to seek an optimal model. The BIC is
 \begin{equation}\label{eq11}
BIC =  - 2\ln {L_{\max }} + k{\kern 1pt} {\kern 1pt} \ln {\kern 1pt} {\kern 1pt} N,
\end{equation}
where ${L_{\max }}$ is the maximum likelihood, $k$ is the number of free parameters of the model, and $N$ is the number of data points.

By comparing BIC in table \ref{tab:1} from fitting for different models to RIQs, we find Model IV has the smallest BIC, which might indicate that the X-ray luminosity of RIQs is not strongly correlated with their radio luminosity. As for RLQs, BIC for Model I, Model II, and Model III are far smaller than Model IV, which implies X-ray luminosity of RLQs is not only connected with optical/UV luminosity but also related to radio luminosity. Moreover, distance modulus for RLQs including FSRLQs and SSRLQs obtained from Model I have smaller dispersion (see fig \ref{fig:1}), relative to Model IV. Meanwhile, for RLQs, the BIC of Model I is less than Model II and Model III, which shows that the luminosity relation ${L_X} \propto L_{UV}^{{\gamma _{uv}}}L_{Radio}^{\gamma _{radio}'}$ for RLQs has some superiority, relative to other models. A possible reason for the luminosity correlations in RLQs is that a fraction of the nuclear X-ray emission is directly or indirectly powered by the radio jet\citep{ Hardcastle2009, Miller2010}, the specific physical mechanism needs to be further understood.

Furthermore, for the fitting results BIC, there is a difference between FSRLQs and SSRLQs for Model I.  The goodness of fit for SSRLQs seems to be better, which implies FSRLQs probably involve more physical processes \citep{Landt2006, Zhu2020}.

\subsection{Analysis of the relation ${L_X} \propto L_{UV{\kern 1pt} {\kern 1pt} }^{{\gamma _{uv}}}L_{Radio{\kern 1pt} {\kern 1pt} }^{\gamma _{radio}^{'}}$  }

\begin{figure}
\includegraphics[width=\linewidth,scale=1.00]{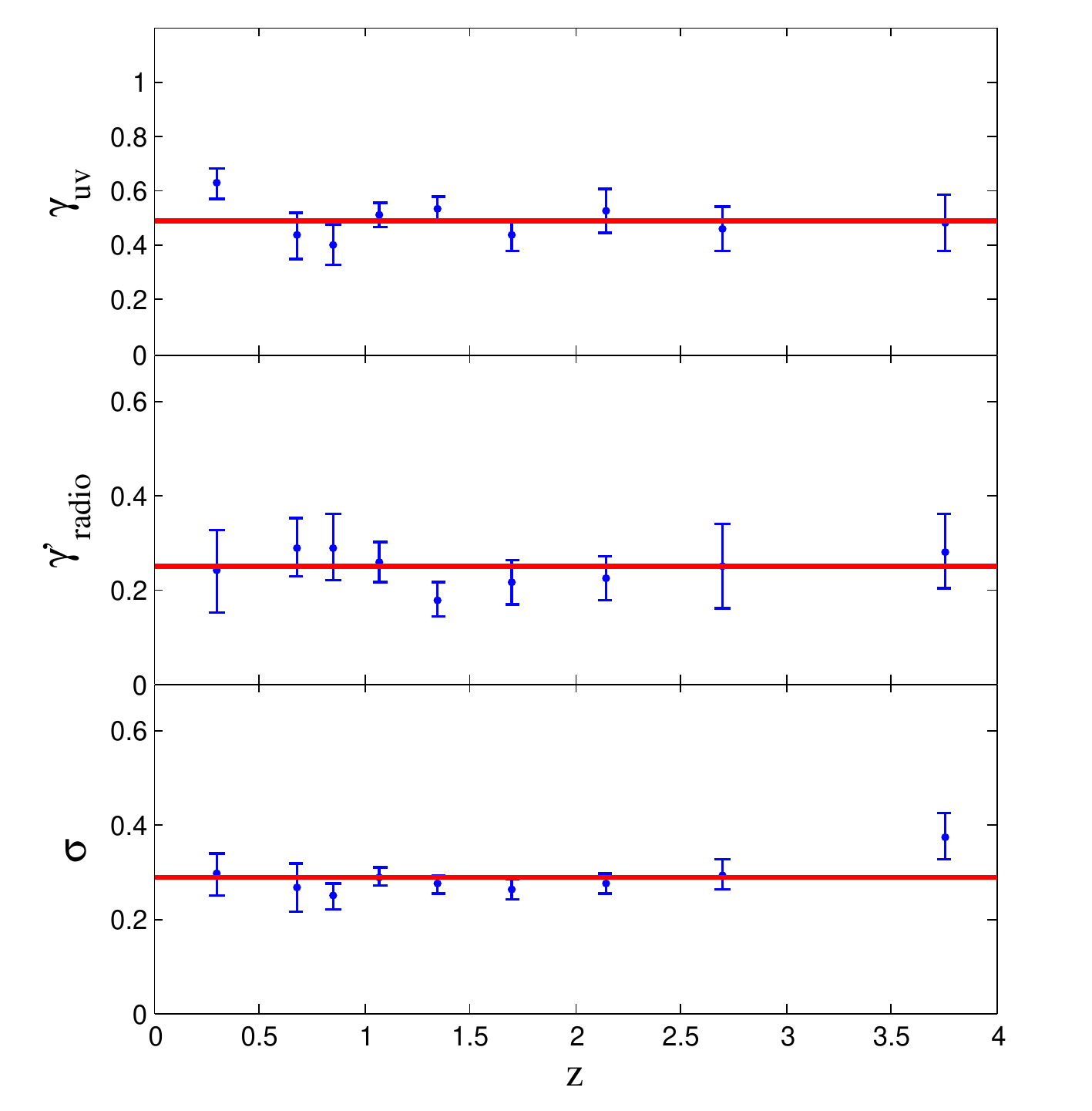}
\caption{${F_X} - {F_{UV}}/{F_{radio}}$ correlation in narrow redshift intervals. Blue points are the fit results of ${\gamma _{UV}}(z),{\kern 1pt} {\kern 1pt} \gamma _{radio}'(z),{\kern 1pt} {\kern 1pt} \sigma (z)$ at different redshift. The horizontal lines show their average values.}
\label{fig:3}
\end{figure}

We divide the RLQs data in several redshit bins, which can be used to check if luminosity relation ${L_X} \propto L_{UV}^{{\gamma _{uv}}}L_{Radio}^{\gamma _{radio}'}$ depend on redshift. The redshift bins satisfy $\Delta \log z = 0.1$. We adopt parametric model \citep{Risaliti2015}
 \begin{equation}\label{eq12}
\log {F_X} = \alpha (z) + {\gamma _{uv}}(z)\log {F_{UV}} + \gamma _{radio}'(z)\log {F_{radio}},
\end{equation}
where $\alpha (z),{\kern 1pt} {\gamma _{uv}}(z),{\kern 1pt} \gamma _{radio}'(z)$ are free parameters. We apply segmented RLQs data to fit ${\kern 1pt} {\gamma _{uv}}(z)$ and ${\kern 1pt} \gamma _{radio}'(z)$  as well as test whether there are a dependency upon redshift. The fit results of ${\gamma _{UV}}(z),{\kern 1pt} {\kern 1pt} \gamma _{radio}'(z),{\kern 1pt} {\kern 1pt} \sigma (z)$ at different redshift are illustrated in fig \ref{fig:3}, it is easy to see that their value do not obviously deviate from the average, which shows there are no obvious evidence for any significant redshift evolution. The average values of parameters are $\left\langle {{\gamma _{uv}}} \right\rangle  = 0.49 \pm 0.165,{\kern 1pt} {\kern 1pt} {\kern 1pt} \left\langle {\gamma _{radio}'} \right\rangle  = 0.248 \pm 0.083.$

\section{The reconstruction of dark energy equation of state w(z)}\label{Sec:5}

Although cosmological constant $\Lambda CDM$ model can be used to effectively explain the accelerating expansion of universe and the cosmic microwave background (CMB) anisotropies \citep{Riess1998,Amanullah2010,Betoule2014,Scolnic2018,Conley2010,Aghanim2020,Hu2002,Spergel2003,Ade2016,Aghanim2016}, the origin and nature of dark energy density and pressure are still unclear.

Dark energy can be studied using two main approaches. The first is to constrain dark energy physical models and attempt to explain the physical origin of its density and pressure \citep{Peebles2003, Ratra1988, Li2004, Maziashvili2007, Amendola2000, Huang2021b, Gao2020}. Understanding the physical nature of dark energy is important for our universe. Whether or not the dark energy is composed of Fermion pairs in a vacuum or Boson pairs, Higgs field, and whether it has weak isospin, which may determine whether it can be observed by experiment. The second is to focus on the properties of dark energy, investigating whether or not its density evolves with time, this can be checked by reconstructing the dark energy equation of state $w(z)$ \citep{Linder2003, Maor2002}, which is independent of physical models. The high redshift observational data can better solve these issues.

 The reconstruction of the equation of state includes parametric and non-parametric methods \citep{Huterer2003,Clarkson2010,Holsclaw2010,Seikel2012,Shafieloo2012,Crittenden2009,Crittenden2012,Zhao2012,Huang2021a}. We use RLQs and SNla to reconstruct $w(z)$ by parametric method assuming X-ray luminosity relation Equation (\ref{eq2}), which can be used for testing the nature of dark energy.

SNla Pantheon sample is the combination of SNIa from the Pan-STARRS1 (PS1), the Sloan Digital Sky Survey (SDSS), SNLS, and various low-z and Hubble Space Telescope samples. There are 279 SNIa provided by PS1 \citep{Scolnic2018}, and SDSS presented 335 SNla \citep{Betoule2014,Gunn2006,Gunn1998,Sako2007,Sako2014}. The rest of Pantheon sample are from the ${\rm{CfA1 - 4}}$, CSP, and Hubble Space Telescope (HST) SN surveys \citep{Amanullah2010,Conley2010}. This extended sample of 1048 SNIa is called the Pantheon sample.

The integral formula of ${D_L} - z$ relation in flat space can be written as
 \begin{equation}\label{eq13}
\begin{array}{l}
{D_L} = \frac{{1 + z}}{{{H_0}}}\int_0^z {d{z'}[{\Omega _m}{{(1 + {z'})}^3}} \\
{\kern 1pt} {\kern 1pt} {\kern 1pt} {\kern 1pt} {\kern 1pt} {\kern 1pt} {\kern 1pt} {\kern 1pt} {\kern 1pt} {\kern 1pt} {\kern 1pt} {\kern 1pt} {\kern 1pt} {\kern 1pt} {\kern 1pt} {\kern 1pt} {\kern 1pt} {\kern 1pt} {\kern 1pt} {\kern 1pt} {\kern 1pt} {\kern 1pt} {\kern 1pt} {\kern 1pt} {\kern 1pt} {\kern 1pt} {\kern 1pt} {\kern 1pt} {\kern 1pt} {\kern 1pt} {\kern 1pt} {\kern 1pt} {\kern 1pt} {\kern 1pt} {\kern 1pt} {\kern 1pt} {\kern 1pt} {\kern 1pt} {\kern 1pt} {\kern 1pt} {\kern 1pt} {\kern 1pt} {\kern 1pt} {\kern 1pt} {\kern 1pt} {\kern 1pt} {\kern 1pt} {\kern 1pt} {\kern 1pt} {\kern 1pt} {\kern 1pt}  + {\Omega _R}{(1 + {z'})^4} + \Omega _{DE}^{(0)}{{\mathop{\rm e}\nolimits} ^{\int_0^{{z'}} {\frac{{1 + w({z^{''}})}}{{1 + {z^{''}}}}d{z^{''}}} }}{]^{ - 1/2}}
\end{array}
\end{equation}
where ${{\Omega _R}}$ is radiation density. ${\Omega _{DE}^{(0)}}$ is the present dark energy density and satisfies $\Omega _{DE}^{(0)} = 1 - {\Omega _m}$ when ignoring ${{\Omega _R}}$, $w(z)$ is dark energy equation of state.
We choose parametric form for $w(z)$
 \begin{equation}\label{eq13}
w(z) = {w_0} + {w_a}\frac{z}{{1 + z}}.
\end{equation}
%and labeling it ${w_0}{w_a}CDM$.
Therefore dark energy density is
 \begin{equation}\label{eq14}
{\Omega _{DE}}(z) = \Omega _{DE}^{(0)}{(1 + z)^{3(1 + {w_0} + {w_a})}}\exp [ - 3{w_a}z/(1 + z)].
\end{equation}

We constrain ${{\rm{w}}_0}{{\rm{w}}_a}{\rm{CDM}}$ model parameters for RLQs and SNla by minimizing $\chi _{Total}^2$, the $\chi _{Total}^2$ is
 \begin{equation}\label{eq15}
\chi _{Total}^2 =  - 2\ln {L^{RLQs}} + \chi _{SN}^2,
\end{equation}
where $ - 2\ln {L^{RLQs}}$ is given by equation (\ref{eq6}), and $\chi _{SN}^2$ can be expressed as
 \begin{equation}\label{eq16}
\chi _{SN}^2 = \Delta {\mu ^T}C_{{\mu _{ob}}}^{ - 1}\Delta \mu ,
\end{equation}
where $\Delta \mu  = \mu  - {\mu _{th}}$. ${C_\mu }$ is the covariance matrix of the distance modulus $\mu $.

\begin{figure}
\includegraphics[width=\linewidth]{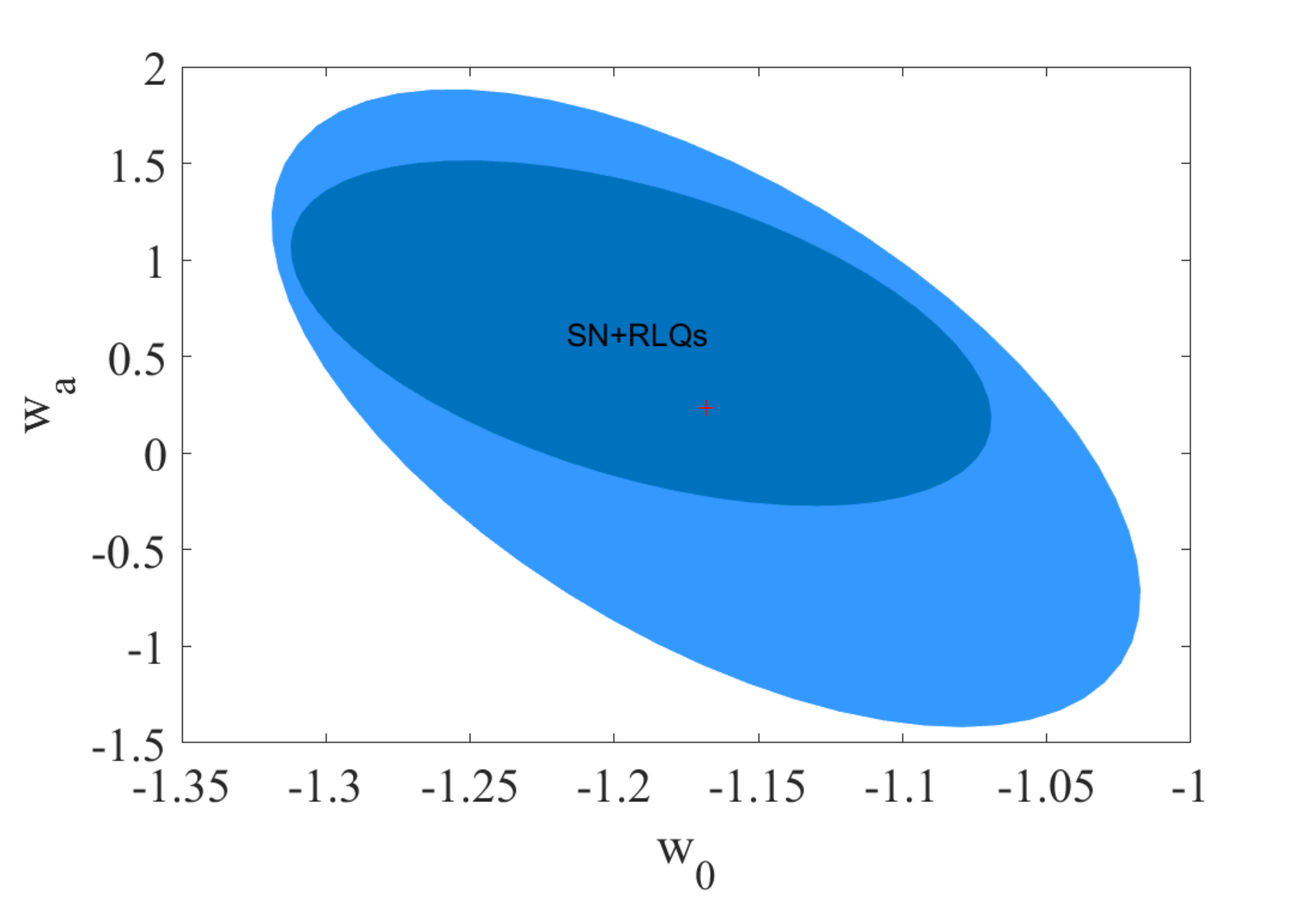}
\caption{$68\%$ and $95\%$ contours for ${w_0}$ and ${w_a}$ from a fit of the X-ray luminosity relation ${L_X} \propto L_{UV}^{{\gamma _{uv}}}L_{Radio}^{\gamma _{radio}'}$ with $\Lambda CDM$ and ${w_0}{w_a}CDM$ model to a combination of SNla and RLQs. The + dot in responding color represents the best fitting values for ${w_0}$, ${w_a}$.}
\label{fig:4}
\end{figure}

We use equation (\ref{eq15}) to fit model parameters, and fit results are shown in table \ref{tab:2}, the results shows ${w_0}{w_a}CDM$ has better goodness of fit than $\Lambda CDM$, and $\Delta \chi _{Total}^2$ is improved by $-3.4$, which indicate $\Lambda CDM$ model is in tension with RLQs at $\sim 1.5\sigma $ . Meanwhile fig \ref{fig:4} illustrates  $68\%$ and $95\%$ contours for ${w_0}$ and ${w_a}$ from a combination of SNla and RLQs, assuming the X-ray luminosity relation ${L_X} \propto L_{UV}^{{\gamma _{uv}}}L_{Radio}^{\gamma _{radio}'}$. The distance modulus and properties of the 571 RLQs are listed in Table \ref{tab:3} (the full data can be downloaded from \href{https://academic.oup.com/mnras/article-abstract/515/1/1358/6595979}{MNRAS} or  \href{http://vizier.u-strasbg.fr/viz-bin/VizieR?-source=J/MNRAS/515/1358}{J/MNRAS/515/1358}).

\begin{table*}
\normalsize
\centerline{}
\caption{The properties of the 571 RLQs, $DM$ are the distance modulus from a fit of the X-ray luminosity relation ${L_X} \propto L_{UV}^{{\gamma _{uv}}}L_{Radio}^{\gamma _{radio}'}$ with $\Lambda CDM$ model, ${\sigma _{DM}}$ are their error.}
\label{tab:3}
\vspace{0.3cm}
\begin{tabular}{cccccccccc}
\hline
$SDSS{\kern 1pt} {\kern 1pt} {\kern 1pt} name $&$z $&${m_i} $&${F_{UV}}$&${F_{radio}}$&${F_X}$&$\log {\kern 1pt} {\kern 1pt} R$&${\alpha _r}$&$DM$&${\sigma _{DM}}$\\
\hline
083946.22+511202.8&4.39 & 18.98& -23.62&	-21.03&	-27.43&	2.46&	0.17&	46.19&	2.014\\
091824.38+063653.3&	4.192&	19.28&	-23.75&	-21.15&	-27.59&	2.47&	0.25&	46.76&	2.014\\
121453.46+300832.7&	3.505&	19.96&	-24.06&	-21.51&	-27.58&	2.43&	-0.77&	45.12&	2.014\\
080928.21+255240.2&	2.853&	20.81&	-24.33&	-21.22&	-27.57&	2.99&	-0.35&	44.7&	2.014\\
122012.04+300250.5&	1.733&	20.13&	-24.06&	-21.14&	-27.29&	2.81&	-0.64&	43.62&	2.014\\

\hline

\end{tabular}

\end{table*}

\section{Summary}\label{Sec:6}

The verification of X-ray luminosity correlation for RIQs and RLQs could make us understand more of their physical mechanism. We obtain a sample of 144 RIQs and 571 RLQs with radio, optical/UV, and X-ray coverage. Firstly, we adopt four parametric methods to test the correlation between X-ray, optical/UV, and radio luminosity. Data suggest that  X-ray luminosity relation ${L_X} \propto L_{UV}^{{\gamma _{uv}}}L_{Radio}^{\gamma _{radio}'}$ is more suitable for RLQs by comparing BIC, which implies X-ray luminosity of RLQs are not only related to optical/UV luminosity but also connected with radio luminosity. Similarly, X-ray luminosity relation ${L_X} \propto L_{UV}^{{\gamma _{uv}}}$ is preferred by RIQs, which indicates that the X-ray luminosity of RIQs is not strongly correlated with their radio luminosity. Meanwhile, we compare the results from FSRLQs and SSRLQs using a fit of X-ray luminosity relation ${L_X} \propto L_{UV}^{{\gamma _{uv}}}L_{Radio}^{\gamma _{radio}'}$, the goodness of fit for SSRLQs seems to be better, which indicates that X-ray luminosity  ${L_X} \propto L_{UV}^{{\gamma _{uv}}}L_{Radio}^{\gamma _{radio}'}$ is more suitable for SSRLQs compared to FSRLQs.

Secondly, We divide the RLQs data into several redshift bins and combine a special model to check if there is a redshift evolution of X-ray luminosity relation ${L_X} \propto L_{UV}^{{\gamma _{uv}}}L_{Radio}^{\gamma _{radio}'}$, the fit results show the model parameters approach to the constant, which implies there is not an obvious redshift evolution for ${L_X} \propto L_{UV}^{{\gamma _{uv}}}L_{Radio}^{\gamma _{radio}'}$.

Finally, we employ a joint of SNla and RLQs to reconstruct the dark energy equation of state, which can be used to test the nature of dark energy. the results show ${w_0}{w_a}CDM$ model is superior to cosmological constant $\Lambda CDM$ model at $\sim 1.5\sigma $.

In the future, we will cross-correlate the SDSS quasar catalogs with the XMM-Newton, Chandra archives, and radio surveys. We expect to obtain several hundred RLQs at high redshift ($z>3$) with multi-wavelength coverage. The high redshift observational data can better test the properties of dark energy, which will dominate the future of the universe, that the universe continue expanding or change from expansion to contraction. It will similarly determine the future of humanity.

\section*{Acknowledgements}
We thank Dr. Ning Chang for his useful discussion, and suggestions for revision.

\section*{DATA AVAILABILITY}
The data underlying this article are available in the article and in its online supplementary material.

\end{document}